\documentclass{iau}
\usepackage{graphicx}

\DeclareRobustCommand{\ion}[2]{%
\relax\ifmmode
\ifx\testbx\f@series
{\mathbf{#1\,\mathsc{#2}}}\else
{\mathrm{#1\,\mathsc{#2}}}\fi
\else\textup{#1\,{\mdseries\textsc{#2}}}%
\fi}

\newcommand{\rev}[1]{{ \textbf}}

\title[JD 11.~~ Searching for outflows in X-ray weak quasars] %% give here short title %%
{Searching for outflows in X-ray weak quasars}

\author[Bartolomeo Trefoloni]   %% give here short author list %%
{Bartolomeo Trefoloni$^1$,$^2$}

\affiliation{
$^1$Dipartimento di Fisica e Astronomia, Universit\`a di Firenze, via G. Sansone 1, 50019 Sesto Fiorentino, Firenze, Italy  \\[\affilskip]

$^2$INAF -- Osservatorio Astrofisico di Arcetri, Largo Enrico Fermi 5, I-50125 Firenze, Italy
\\ {\tt email: bartolomeo.trefoloni@unifi.it}}

\pubyear{2023}
\volume{378}  %% insert here IAU Symposium No.
\pagerange{119--126}
\jname{Black hole winds at all scales}
\editors{A.C. Editor, B.D. Editor \& C.E. Editor, eds.}
\begin{document}

\maketitle

\begin{abstract}
The connection between X-ray weakness and powerful X-ray outflows is both expected in a scenario where outflows are connected with radiation-driven winds, and observed in several sources, both in the local Universe and at high redshift. Here I present the first results of a new study of this possible connection based on a search for SDSS quasars with weak X-ray emission in serendipitous XMM-Newton observations. The selected objects have a "normal" optical/UV blue continuum, but a flat and extraordinarily weak X-ray spectrum. The availability of rest-frame optical/UV spectra allows to check for the signature of outflows in the absorption lines and/or in the profiles of the emission lines. This method could reveal the
presence of a population of so-far overlooked outflowing quasars and confirm the
connection between winds and X-ray weakness in quasars.

%% add here a maximum of 10 keywords, to be taken form the file <Keywords.txt>
\end{abstract}

\section{Introduction}
Quasars are the most luminous persistent sources in the Universe, and there is growing evidence of their relevance for the observed properties of their host galaxies. It is general thought that an accretion disc around super-massive black holes (BH) produces their main contribution in the optical/UV (e.g.,  \cite[Salpeter 1964, Lynden-Bell 1969, Czerny \& Elvis 1987]{salpeter_64, lynden_bell_69, czerny_elvis_87}), while the X-ray emission is likely a product of the so-called ``corona'' (e.g., \cite[Sunyaev \& Titarchuck 1980, Haardt \& Maraschi 1993]{sunyaev_titarchuck_80, haardt_maraschi_93}), where UV disc photons are boosted via inverse Compton scattering. These two emission features are closely related (e.g. \cite[Lusso et al. 2020]{lusso_20} and references therein), although the physics underlying such interplay is still poorly understood, and their interplay could in principle vary with the accretion parameters. At high accretion rate, for instance, the hypotheses of geometrically thin and optically thick disc (\cite[Shakura \& Sunyaev 1973]{shakura_sunyaev_73}) could break down, with the thickening of the disc (\cite[Abramowicz et al. 1988, Chen \& Wang 2004, Jiang et al. 2014]{abramowicz_88, chen_wang_04, jiang_19}).

In addition, powerful accretion-disc winds directly related to the nuclear activity (e.g., \cite[Proga 2005]{proga_05}) could significantly alter the standard accretion process. Near Eddington sources are more likely to host of such outflows (\cite[Zubovas \& King 2013, Nardini et al. 2015]{zubovas_king_13, nardini_15}), which could underlie the observed relations between the black hole mass ($M_{\rm BH}$) and the galaxy properties (e.g., the $M_{\rm BH}-\sigma$ relation; \cite[Ferrarese \& Merrit 2000, Gebhardt et al.  2000]{ferrarese_merrit_00, gebhardt_00}), although it is not yet clear whether and how AGN-driven outflows can affect their host galaxies. 

At the same time, several samples of highly accreting quasars sharing quite homogeneous UV properties have recently shown an enhanced fraction of objects whose X-ray spectra are relatively flat and underluminous (by factors of $>3-10$) with respect to the expectations from  the $L_{\rm X}-L_{\rm UV}$ relation (e.g., \cite[Luo et al. 2015, Nardini et al. 2019, Zappacosta et al. 2020, Laurenti et al. 2022]{luo_15, nardini_19, zappacosta_20, laurenti_22}). In many cases any clear evidence for absorption has been revealed by the spectral analysis. 

\section{The parent sample}
With the aim of expanding the $L_{\rm X}-L_{\rm UV}$ relation sample in the high luminosity and high redshift tail of the SDSS distribution, our group obtained dedicated XMM-Newton observations of 30 luminous quasars between 3.0$\leq z \leq$3.3. This sample was selected in order to present a high degree of homogeneity in terms of UV properties, while satisfying the customary requirements of the $L_{\rm X}-L_{\rm UV}$ relation sample (e.g. \cite[Risaliti \& Lusso 2019]{risaliti_lusso_19}). This implied the exclusion of known broad absorption lines (BALs) objects, as well as Radio--loud ($F_{\nu,6cm}/F_{\nu,2500\AA}>10$) and extincted ($E(B-V)>$0.1) ones. Despite sharing such optical/UV properties, the sample presented a diverse behaviour in terms of X-ray characteristics (\cite[Nardini et al. 2019]{nardini_19}).
About two thirds of the sample show X-ray luminosities in agreement with the expectations from the $L_{\rm X}-L_{\rm UV}$ relation ($N$ quasars), and a mean photon index of $\Gamma_{\rm X}$\,$\sim$\,1.85, fully consistent with typical quasars at lower redshift, luminosity, and $M_{\rm BH}$ (e.g. \cite[Bianchi et al. 2009]{bianchi_09}). Their luminosity in the 2--10 keV band spans between $4.5\times 10^{44}\leq L_{2-10\,\rm keV} \leq 7.2\times 10^{45}$ erg/s, being one of the most X-ray luminous samples of radio-quiet quasars ever observed. On the other hand, one third of the sources proved underluminous by factors $\sim$\,3--10 ($W$ quasars). X-ray absorption at the redshift of the source is not generally statistically required by the fits of the X-ray spectra. Notwithstanding the poor quality of the data in some of cases, which leaves room for minor absorption, column densities $N_{\rm H}(z) > 3 \times 10^{22}$ cm$^{-2}$ can be ruled out.

We continued our investigation about the spectral differences between $N$ and $W$ quasars by performing a detailed analysis of the SDSS archival spectra for the whole sample (\cite[Lusso et al. 2021]{lusso_21}). These spectra cover the rest-frame region roughly between 900-2200 \AA\ including the \ion{C}{iv}$\lambda$1549 emission line. We focused on such emission feature to characterize its line properties (e.g., equivalent width, EW; line peak velocity, $v_{peak}$) and underlying UV continuum slope as a function of the X-ray photon index and 2--10 keV flux. We found that the composite spectrum of X-ray weak quasars is flatter ($\alpha_{\lambda}\sim -0.6$) than the one of X-ray normal quasars ($\alpha_{\lambda}\sim -1.5$). The \ion{C}{iv} emission line is in general fainter in $W$ quasars, but we did not report a strong blueshift (600--800 km/s) in both stacks. Such emission feature appears to be broader in the $W$ composite spectrum (FWHM $\simeq$ 10,000 km/s) than in the $N$ one ($\simeq$\,7,000 km/s), but this result is not dissimilar from other literature samples at similar redshifts (e.g. \cite[Shen et al. 2011]{shen_11}) and luminosities (e.g. \cite[Vietri et al. 2018]{vietri_18}). We also added the sample from \cite[Timlin et al. 2020]{timlin_20}, adopting our selection criteria, with the aim to expand the dynamical range of the parameters of interest. In such a way, we confirmed the statistically significant trends of \ion{C}{iv} $v_{peak}$ and EW with UV luminosity at 2500 \AA\ for both $W$ and $N$, as well as the correlation between X-ray weakness and the EW of \ion{C}{iv}. In addition, a statistically significant correlation between the hard X-ray flux and the integrated \ion{C}{iv} flux for X-ray normal quasars was found. This relation spans more than three decades in \ion{C}{iv} and two in terms of X-ray luminosity. X-ray weak quasars drop from the bulk of the relation by more than 0.5 dex. 

Lastly, our group was awarded new dedicated observations at the Large Binocular Telescope (LBT) in the $zJ$ and $K_S$ bands corresponding to the rest frame \ion{Fe}{ii}\textsubscript{UV}-\ion{Mg}{ii}$\lambda$2798 and the \ion{H}{$\beta$}-[\ion{O}{iii}] emission regions (\cite[Trefoloni et al. 2023]{trefoloni_23}). These new data allowed to put tighter constraints on BH masses by taking advantage of \ion{H}{$\beta$} and \ion{Mg}{ii}$\lambda$2798 single-epoch calibrations (\cite[Shen et al. 2011, Bongiorno et al. 2014]{shen_11, bongiorno_14}), more reliable than the \ion{C}{iv} ones, so far available. We found that the sample is on average made of quasars accreting near the Eddington limit $<\lambda_{Edd}>=1.2$ ($\lambda_{Edd}$ = $L_{bol}/L_{Edd}$, being $L_{bol}$ and $L_{Edd}$ respectivelythe bolometric and the Eddington luminosity), as expected for such extremely luminous objects. By fixing the slope of the continuum in order to match the one of the bluer SDSS spectral counterpart, a trend has emerged with $W$ objects generally exhibiting more prominent \ion{Fe}{ii}\textsubscript{UV} emission and, in turn, higher \ion{Fe}{ii}/\ion{Mg}{ii} ratios with respect to $N$ ones. Such feature ratio could be enhanced in case of wind-related microturbulence and shocks in outflows (e.g. \cite[Baldwin et al. 2004, Sameshima et al. 2017, Temple et al. 2020]{baldwin_04, sameshima_17, temple_20}). The EW\,[\ion{O}{iii}] is generally low ($<$20 \AA) in all but one of our objects. Inclination effects are not expected to play a dominant role, since sources observed under a highly inclined line of sight with respect to the axis of the disc usually display EW\,[\ion{O}{iii}] $\gtrsim$ 30 \AA. We also observed a lower EW of several emission lines (\ion{C}{iv}, \ion{H}{$\beta$}, [\ion{O}{iii}]) in X-ray weak quasars, and we argue that this effect could be related to the decrease of EUV photons responsible for the line production. Indeed, the $L$\textsubscript{[\ion{O}{iii}]} of the both X-ray weak and normal quasars is consistent with the high-luminosity extrapolation of the $L$\textsubscript{[\ion{O}{iii}]}--$L_{\rm{2-10\,keV}}$ relation from other samples in literature suggesting a common origin for the intrinsically low X-ray and [\ion{O}{iii}] emission, rather than an inclination/obscuration scenario.

We argue that, even though we do not diffusely observe unequivocal outflow footprints (e.g. large \ion{C}{iv} and/or [\ion{O}{iii}] bluewings, BALs) in X-ray weak quasars, the presence of a mostly equatorial disc wind could explain the observational features reported so far (Figure 1). Part of the UV radiation would not be reprocessed in the X-ray corona causing intrinsic (i.e., not due to absorption) X-ray weakness. The lack of distinctive outflow signatures could be due to the modest inclination of the line of sight to the disc, consistent with the observed EW\,[\ion{O}{iii}] values that we find and with the huge bolometric luminosities. A higher \ion{Fe}{ii}\textsubscript{UV} emission in X-ray weak quasars could be produced by regions of microturbulence and shocks at the interface between the wind and the BLR gas. In addition, the effect of the change in the local mass accretion rate could dramatically alter the EUV SED, and ultimately determine the observed difference in terms of emission line strength (\ion{C}{iv}$\lambda$1549, \ion{H}{$\beta$}, [\ion{O}{iii}]). The incidence of this mechanism could be higher in samples of highly accreting sources.

\begin{figure}[h!]
  \centering
  \begin{minipage}[b]{0.49\textwidth}
    %\vspace{10mm}
    \includegraphics[scale=0.25]{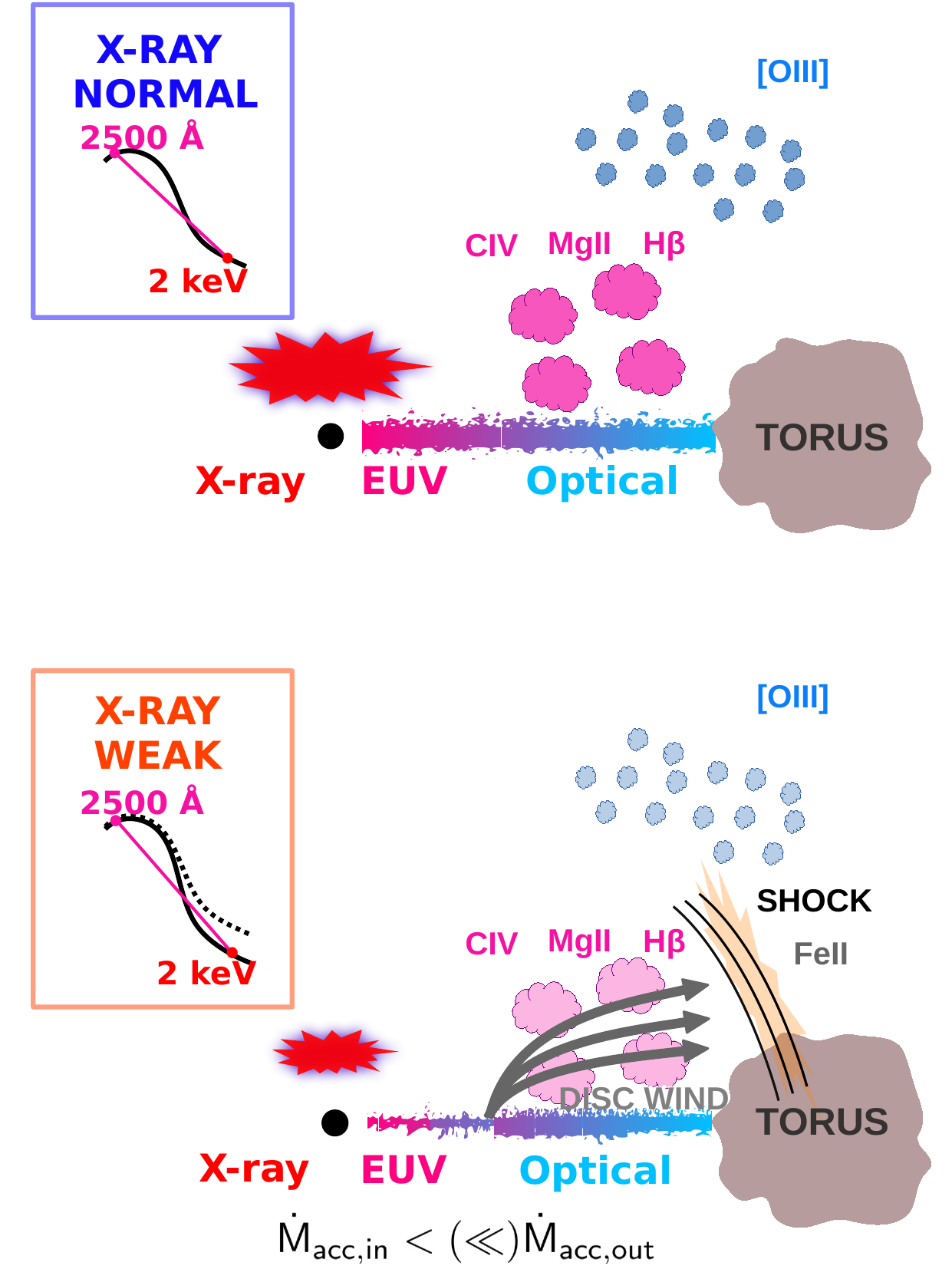}
    \caption{A sketch describing the observational features of X-ray normal (top) and weak (bottom) quasars.}
  \end{minipage}
  \hfill
  \begin{minipage}[b]{0.49\textwidth}
    \includegraphics[width=1\textwidth]{proced_sample.pdf}
    \caption{The $L_X-L_{UV}$ sample from \cite[Lusso et al. 2020]{lusso_20}, here used as a benchmark to select X-ray weak sources. In red are depicted the $W$ sources. Large red stars represent the pilot sample, squares the z$\sim$3 sample.}
  \end{minipage}
\end{figure}

\section{A new pilot sample}

Prompted by such results, we explored the possible correlation between the \textit{exotic} coronal state characterized by an underluminous and flat X-ray spectra and optical/UV outflow signatures. To this purpose, we designed a new pilot sample, made of less luminous analogues of the $W$ quasars, in order to investigate the tentative relation between the flat and weak X-ray emission and the possible outflowing phase. We required the sample to be X-ray underluminous by factors $>$3, while still presenting good X-ray and UV/optical quality (SN$\gtrsim$5) in order to be able to detect absorption and/or outflows, if present. We also required a photometric photon index $\Gamma + \delta \Gamma < 1.3$ (see \cite[Lusso et al. 2020]{lusso_20} for more details). Since for the optical/UV side of the emission we relied on archival SDSS data, we restricted the analysis to the redshift intervals z$\leq$0.8 and 1.8$\leq$z$\leq$2.7 where respectively the \ion{H}{$\beta$}-[\ion{O}{iii}]$\lambda \lambda$4959,5007 complex and the \ion{C}{iv}$\lambda$1549-\ion{Mg}{ii}$\lambda$2798 fall. Adopting such redshift intervals, we have also been able to retain at least one reliable BH mass virial estimator to provide an estimate of the accretion parameters ($M_{BH}$, $\lambda_{Edd}$).
These criteria yielded a sample of 16 objects, 8 with UV spectra and 8 with optical spectra. We performed a detailed spectral analysis on X-ray and optical/UV spectra. In the X-rays we adopted a power law spectrum and included both the Galactic and the local absorption with the last one turning out to be not statistically significant in general. This meant that the observed weakness was \textit{intrinsic} rather than due to obscuration.
To perform the spectral decomposition of the optical/UV spectra, we assumed a power-law continuum and different emission line profiles (Gaussian, Lorentzian). We also allowed for a blue component in both \ion{C}{iv}$\lambda$1549 and [\ion{O}{iii}]$\lambda \lambda$4959,5007 and included templates for the \ion{Fe}{ii}\textsubscript{UV} and \ion{Fe}{ii}\textsubscript{opt} pseudo-continua. Three sources out of eight with optical spectra display a prominent [\ion{O}{iii}] bluewing with offset velocity in excess of $\sim$200 km/s and broad $\gtrsim$850 km/s profiles, likely produced in galactic-scale outflows. Half of the sources with UV spectra either show blueshifted ($\gtrsim$ 3000 km/s)  or broad ($\gtrsim$2000 km/s) absorption components in the \ion{C}{iv} profile. As a rough comparison only 2\% and 3\% of the sources reported in the latest SDSS catalogue (\cite[Wu \& Shen 2022]{wu_shen_22}) exhibit blueshift in excess of 200 and 2000 km/s for the [\ion{O}{iii}] and \ion{C}{iv} emission lines respectively\footnote{Not being the parameters of the blue component available in such catalogue, these estimates are based on the blueshift of the single broad component for the \ion{C}{iv} and on the difference between the global profile and the core peak wavelength for the [\ion{O}{iii}].}.
As a side note, we should also mention that the X-ray and UV/optical spectral data have not been gathered simultaneously, thus time shift effects could also dilute or even obliterate actual correlations.

Although these preliminary results have been drawn from a relatively small pilot sample, a unifying picture relating the nuclear X-ray emission and the outer multi-scale outflow phases is emerging (e.g. \cite[Zubovas \& King 2014, Cicone et al. 2014, Gaspari et al. 2020]{zubovas_king_14, cicone_18, gaspari_20}). In this framework, the weak phase of the corona could play a key role to set the environmental conditions to launch fast winds. Otherwise an efficient X-ray emission could overionize the nuclear medium, making wind line-driving unfeasible. 

The results presented in these proceedings only represent a glimpse of what we believe could bridge the observed state of X-ray underluminosity to the possibly related BLR emission properties. Such multi-wavelength approach yields the potential to pave the way for a more complete description and understanding of the interplay between these two bands of the panchromatic quasar emission.

\end{document}